\documentclass[twocolumn,showpacs,preprintnumbers,pra]{revtex4}
\usepackage{amssymb}
\usepackage{amsmath}
\usepackage{graphicx}
\usepackage{dcolumn}
\usepackage{bm}

\begin{document}

\title{Measurable genuine tripartite entanglement of ($2\otimes 2\otimes n$%
)-dimensional quantum states via  only two simultaneous  copies}
\author{Chang-shui Yu}
\email{quaninformation@sina.com;ycs@dlut.edu.cn}
\author{Bao-qing Guo}
\author{Si-ren Yang}
\affiliation{School of Physics and Optoelectronic Technology, Dalian University of
Technology, Dalian 116024, China }
\date{\today }

\begin{abstract}
Usually, the three-tangle of a tripartite pure state of qubits can be directly measured 
with the simultaneous preparation of  a not-less-than-four-fold copy of the state. We show that the exact genuine tripartite entanglement for ($2\otimes
2\otimes n$)-dimensional pure quantum states can be measured in a similar manner, provided that only  two simultaneous copies of the state are available.  Lower bounds  are also proposed for more convenient experimental operations. As an example, a comprehensive demonstration of the scheme is
provided for the three-tangle of a three-qubit state.
\end{abstract}

\pacs{03.67.Mn, 42.50.-p}
\maketitle

\section{Introduction}
 Quantum entanglement is the combination of quantum
superposition and the tensor product structure of quantum state space. It is
one of the most fundamental features of quantum mechanics which distinguishes
the quantum from classical world , while it serves as an important physical
resource in quantum information processing tasks. In past decades,
quantification of entanglement, one of the key subjects in entanglement
theory, has attracted much interest and a lot of remarkable entanglement
measures have been proposed \cite{sum}. However, quantum entanglement, in general, does
not correspond to an observable due to the unphysical operations such as the
complex conjugation for concurrence \cite{conc} and the partial transpose for
negativity \cite{neg1,neg2}. This means that entanglement can not be directly measured
in experiment. So the usual method to measuring entanglement is 
reconstructing the density matrix to be considered by the state tomography \cite{tom1,tom2,tom3}
which is fit for small systems. As an effective tool of detecting
entanglement, the entanglement witness \cite{wit} usually depends on the state, which
implies some prior knowledge about the state. Even though
some methods have been developed in order to overcome the mentioned
shortcomings \cite{coh,lo,eig1,eig2,purity}, an important step is taken by the reformulation of 
pure-state concurrence of qubits in terms of a series of projectors on the two-fold copy of the state \cite{copym1,copym2}.  This is also a direct
motivation for both the latter experimental realization \cite{prae,ne} and the theoretic
progress on the measurable concurrence for mixed states \cite{c1,c2,c3}, measurable
geometric discord of two qubits \cite {disc,disc2}, and measurable three-tangle of pure states \cite{m3t}.
Of course, the two-fold (or multiple-fold) copy should be understood as a source  producing identical states
to the one for which we want to quantify its entanglement, which should be distinguished from the quantum non-cloning theorem \cite{cloning}. Although a not-less-than-fourfold copy of a state for this entanglement makes the
direct measurements possible, it simultaneously challenges the practical
realization in experiment.

In this paper, we present a scheme to directly measure the genuine
tripartite pure-state entanglement not only for qubit systems but also for $%
(2\otimes 2\otimes n)$-dimensional systems. The distinct advantage of this
scheme is as follows. (1) \emph{Only two simultaneous  copies}: only a two-fold copy of the
tripartite quantum state is required compared with the previous four-fold 
copy; (2) \emph{a few projectors}: $2n$ single-party plus two two-party rank-one
projectors are  much less than $(16n^2-1)$ for the state tomography of a $(2\otimes 2\otimes n)$-dimensional state, which is the common advantage of all the related schemes, but only two two-party projectors are needed in
contrast to six for  three-qubit states in the previous scheme \cite{m3t}; (3) \emph{local
operations}: the projective measurements are performed locally, which is the
same as the previous schemes. As an example, we give a detailed
demonstration and analysis on directly measuring the polarization entanglement
of three photons in the frame of linear optics. This shows the feasibility
of our scheme. In addition, the lower bounds of the genuine tripartite entanglement which is especially sufficient  for detecting this type of entanglement  are 
also provided for less adjustments of practical operations, but the cost, besides the lower bound, is that more projective measurement outcomes are needed. Finally, the influence of the imperfect experiment on our scheme is also discussed.

\section{The genuine tripartite entanglement}
Unlike bipartite entanglement,
multipartite entanglement can be divided into many inequivalent entanglement
classes. For example, three qubits can be entangled in two ways \cite{3c} and four
qubits can be entangled in nine ways \cite{4c}. Therefore, usually a single scalar can
only effectively characterize the entanglement of a single class or for some
particular purposes. Even though multipartite entanglement of several
quantum states has been well classified, three-tangle was first presented by
Coffman, \textit{et al.} \cite{3t} and is the most remarkable and widely
accepted entanglement monotone for a general state (instead of the states of
given class) to quantify the Greenberger-Horne-Zeilinger (GHZ) type
entanglement of qubits. GHZ type entanglement describes genuine tripartite
inseparability. It is distinguished from its opposite tripartite entanglement
class (W type entanglement) that lies in the different robustness of the residual
two-qubit entanglement against losing the third qubit. In fact, GHZ type
entanglement can also be understood by the maximal extra average two-qubit
entanglement induced by measurements on the third qubit with classical
communication \cite{2nn}. Furthermore, 3-tangle can be naturally
generalized to a $(2\otimes 2\otimes n)$-dimensional quantum state in terms
of concurrence and localizable concurrence. This can be explicitly given as
follows. A tripartite quantum pure state of qubits defined in the $%
(2\otimes 2\otimes n)$-dimensional Hilbert space $\mathcal{H}=\mathcal{H}%
_{A}\otimes \mathcal{H}_{B}\otimes \mathcal{H}_{C}$ can be written in a
computational basis as 
\begin{equation}
\left\vert \psi \right\rangle
_{ABC}=\sum_{i,j=0}^{1}\sum_{k=0}^{n-1}a_{ijk}\left\vert i\right\rangle
_{A}\left\vert j\right\rangle _{B}\left\vert k\right\rangle _{C}.  \label{s}
\end{equation}%
The reduced density matrix $\rho _{AB}=Tr_{C}\left( \left\vert \psi
\right\rangle _{ABC}\left\langle \psi \right\vert \right) $. Considering the
operations on qudit C, the maximal average concurrence of qubits A and B is
characterized by the localizable concurrence \cite{coa0,coa} which is given by 
\begin{equation}
C_{a}\left( \left\vert \psi \right\rangle _{ABC}\right)
=\sum\limits_{i=1}^{4}\lambda _{i}  \label{lc}
\end{equation}%
and the minimal average concurrence is given by the concurrence \cite{conc} of $\rho
_{AB}$, that is, 
\begin{equation}
C\left( \rho _{AB}\right) =\max \{0,2\lambda _{1}-C_{a}\left( \left\vert
\psi \right\rangle _{ABC}\right) \},  \label{c}
\end{equation}%
where $\lambda _{i}$ denotes the square root of the eigenvalues of the matrix,
\begin{equation}
R=\rho _{AB}\left( \sigma _{y}\otimes \sigma _{y}\right) \rho _{AB}^{\ast
}\left( \sigma _{y}\otimes \sigma _{y}\right)  \label{m}
\end{equation}%
in decreasing order with $\sigma _{y}=\left( 
\begin{array}{cc}
0 & -i \\ 
i & 0%
\end{array}%
\right) $. Thus the genuine tripartite entanglement monotone can be defined
by \cite{2nn}%
\begin{equation}
\tau \left( \left\vert \psi \right\rangle _{ABC}\right) =\sqrt{%
C_{a}^{2}\left( \left\vert \psi \right\rangle _{ABC}\right) -C^{2}\left(
\rho _{AB}\right) }.  \label{3e}
\end{equation}%
It is obvious that $\tau ^{2}\left( \left\vert \psi \right\rangle
_{ABC}\right) $ will become three-tangle for $n=2$. Our first result  shows
that $\tau \left( \left\vert \psi \right\rangle _{ABC}\right) $ can be
directly measured in experiment provided that only two-fold copy of $%
\left\vert \psi \right\rangle _{ABC}$ is available. Our second result
shows that the lower bound of $\tau \left( \left\vert \psi \right\rangle
_{ABC}\right) $ can also be directly measured under the same condition with
much simpler practical operations. 
\begin{figure}[tbp]
\includegraphics[width=4cm]{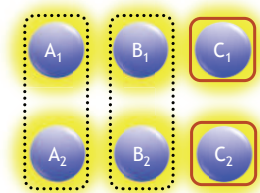}
\caption{(color online) Illustration of the scheme. The three balls in the
yellow shades denote a copy of $\left\vert \protect\psi \right\rangle
_{ABC} $ with the yellow shades representing the potential existence of
quantum correlation. The dotted frame means a projective measurement $P_{-}$
performed on the two particles inside and the solid frame denotes a
single-qubit projective measurement $P$.}
\label{1}
\end{figure}

\section{The measurable tripartite entanglement with two-fold copy}
It can be
found that the matrix $R$ given in Eq. (\ref{m}) is the key to obtaining the
genuine tripartite entanglement monotone $\tau \left( \left\vert \psi
\right\rangle _{ABC}\right) $. Next we will construct another measurable
matrix  that can extract all of the useful information related to $\tau
\left( \left\vert \psi \right\rangle _{ABC}\right) $ from the matrix $R$.

Considering a set of orthonormal basis $\{\left\vert a_{k}\right\rangle \}$
in $\mathcal{H}_{C}$, $\left\vert \psi \right\rangle _{ABC}$ can always be
rewritten by $\left\vert \psi \right\rangle
_{ABC}=\sum_{k=0}^{n-1}\left\vert \varphi _{k}\right\rangle _{AB}\left\vert
a_{k}\right\rangle _{C}$, with $\left\vert \varphi _{k}\right\rangle _{AB}$
denoting the bipartite pure state without normalization. Based on these $\left\vert\varphi_k\right\rangle$, one can easily construct the following symmetric
matrix $M$:
\begin{equation}
M_{ij}=\left\langle \varphi _{i}^{\ast }\right\vert _{AB}\left( \sigma
_{y}\otimes \sigma _{y}\right) \left\vert \varphi _{j}\right\rangle _{AB}.
\label{mm}
\end{equation}%
Thus one can find that the following lemma holds.

\textbf{Lemma 1.-}\textit{The set of the nonzero singular values of the matrix }$M$ \textit{is completely equal to}%
\textit{the set of  the  square root of the eigenvalues of the matrix }$R$.

\textbf{Proof. }At first, we note that the reduced density matrix $\rho
_{AB} $ can be written as $\rho _{AB}=$\textbf{\ }$\sum_{k=0}^{n-1}\left%
\vert \varphi _{k}\right\rangle _{AB}\left\langle \varphi _{k}\right\vert $.
So we can construct an \textit{n}-dimensional matrix $\Psi $ such that 
\begin{equation}
\Psi =\left[ \left\vert \varphi _{0}\right\rangle ,\left\vert \varphi
_{1}\right\rangle ,\left\vert \varphi _{2}\right\rangle ,\cdot \cdot \cdot
,\left\vert \varphi _{n-1}\right\rangle \right] ,  \label{psi}
\end{equation}%
where we have omitted the subscripts $(AB)$. Thus it can be easily found that $%
\rho _{AB}=$ $\Psi \Psi ^{\dag }$, and $M$ can be rewritten as $M=\Psi
^{T}\left( \sigma _{y}\otimes \sigma _{y}\right) \Psi $. In order to find
the singular values, we will have to calculate the eigenvalues of $%
MM^{\dagger }=\Psi ^{T}\left( \sigma _{y}\otimes \sigma _{y}\right) \Psi
\Psi ^{\dagger }\left( \sigma _{y}\otimes \sigma _{y}\right) \Psi ^{\ast }$.
It is obvious that $MM^{\dagger }$ has the same eigenvalue set as the matrix 
$\Psi \Psi ^{\dagger }\left( \sigma _{y}\otimes \sigma _{y}\right) \Psi
^{\ast }\Psi ^{T}\left( \sigma _{y}\otimes \sigma _{y}\right) =\rho
_{AB}\left( \sigma _{y}\otimes \sigma _{y}\right) \rho _{AB}^{\ast }\left(
\sigma _{y}\otimes \sigma _{y}\right) $, i.e., the matrix $R$. This finishes
the proof.\hfill $\blacksquare $

On the basis of Lemma 1, one can draw the conclusion that the genuine
tripartite entanglement $\tau \left( \left\vert \psi \right\rangle
_{ABC}\right) $ will be completely determined once the matrix $M$ is known.
Considering the two-fold copy of $\left\vert \psi \right\rangle _{ABC}$,
i.e., $\left\vert \psi \right\rangle _{A_{1}B_{1}C_{1}}\otimes \left\vert
\psi \right\rangle _{A_{2}B_{2}C_{2}}$ in the Hilbert space $\mathcal{H}_{1}%
\mathcal{\otimes H}_{2}=\left( \mathcal{H}_{A_{1}}\otimes \mathcal{H}%
_{B_{1}}\otimes \mathcal{H}_{C_{1}}\right) \mathcal{\otimes }$ $\left( 
\mathcal{H}_{A_{2}}\otimes \mathcal{H}_{B_{2}}\otimes \mathcal{H}%
_{C_{2}}\right) $, one can define the projectors in the anti-symmetric
subspace $\mathcal{H}_{i_{m}}\wedge \mathcal{H}_{i_{n}}^{\prime }$ of $%
\mathcal{H}_{i_{m}}\otimes \mathcal{H}_{i_{n}}^{\prime }$ as 
\begin{equation}
P_{-}^{(i_{m}i_{n})}=\left\vert \Psi _{i_{m}i_{n}}^{-}\right\rangle
\left\langle \Psi _{i_{m}i_{n}}^{-}\right\vert  \label{pro}
\end{equation}%
with $\left\vert \Psi _{i_{m}i_{n}}^{-}\right\rangle =\frac{1}{\sqrt{2}}%
\left( \left\vert 0\right\rangle _{i_{m}}\left\vert 1\right\rangle
_{i_{n}}-\left\vert 1\right\rangle _{i_{m}}\left\vert 0\right\rangle
_{i_{n}}\right) $ written in the computational basis$,$ where $i=A,B$
corresponds to the subsystem A and B, and $m,n=1,2,$ marks the different
copies of $\left\vert \psi \right\rangle _{ABC}$. Thus we can arrive at
another important lemma.

\textbf{Lemma 2}.-\textit{The entries of }$M$ \textit{can be given, subject
to the two-fold copy of }$\left\vert \psi \right\rangle _{ABC}$\textit{, by} 
\begin{gather}
M_{ij}=2\left( \left\langle \Psi _{A_{1}A_{2}}^{-}\right\vert \left\langle
\Psi _{B_{1}B_{2}}^{-}\right\vert \right) \left( \left\vert \varphi
_{i}\right\rangle _{A_{1}B_{1}}\left\vert \varphi _{j}\right\rangle
_{A_{2}B_{2}}\right)  \label{main1} \\
=2\left( \left\langle \Psi _{A_{1}A_{2}}^{-}\right\vert \left\langle \Psi
_{B_{1}B_{2}}^{-}\right\vert \left\langle i_{C_{1}}\right\vert \left\langle
j_{C_{2}}\right\vert \right) \left( \left\vert \psi \right\rangle
_{A_{1}B_{1}C_{1}}\left\vert \psi \right\rangle _{A_{2}B_{2}C_{2}}\right) ,
\label{main2}
\end{gather}%
\textit{where }$\left\vert i_{C_{1}}\right\rangle $\textit{\ and }$%
\left\vert i_{C_{2}}\right\rangle $\textit{\ are the basis in }$\mathcal{H}%
_{C_{1}}$\textit{\ and }$\mathcal{H}_{C_{2}}$\textit{, respectively}.

\textbf{Proof.} Expand $\sigma _{y}\otimes \sigma _{y}$ in the computational
basis, and we have $\sigma _{y}^{A}\otimes \sigma _{y}^{B}=-\left\vert
00\right\rangle _{AB}\left\langle 11\right\vert +\left\vert 01\right\rangle
_{AB}\left\langle 10\right\vert +\left\vert 10\right\rangle
_{AB}\left\langle 01\right\vert -\left\vert 11\right\rangle
_{AB}\left\langle 00\right\vert $, where we use the indices A and B to mark
different action objects. Therefore, $M_{ij}$ can be rewritten as 
\begin{gather}
M_{ij}=\left\langle \varphi _{i}^{\ast }\right\vert _{AB}\left( \sigma
_{y}\otimes \sigma _{y}\right) \left\vert \varphi _{j}\right\rangle _{AB} \\
=\left[ -\left\langle 00\right\vert _{A_{1}B_{1}}\left\langle 11\right\vert
_{A_{2}B_{2}}\right. +\left\langle 01\right\vert _{A_{1}B_{1}}\left\langle
10\right\vert _{A_{2}B_{2}}  \notag \\
+\left\langle 10\right\vert _{A_{1}B_{1}}\left\langle 01\right\vert
_{A_{2}B_{2}}-\left. \left\langle 11\right\vert _{A_{1}B_{1}}\left\langle
00\right\vert _{A_{2}B_{2}}\right] \left\vert \varphi _{i}\right\rangle
_{A_{1}B_{1}}\left\vert \varphi _{j}\right\rangle _{A_{2}B_{2}}  \notag \\
=2\left( \left\langle \Psi _{A_{1}A_{2}}^{-}\right\vert \left\langle \Psi
_{B_{1}B_{2}}^{-}\right\vert \right) \left( \left\vert \varphi
_{i}\right\rangle _{A_{1}B_{1}}\left\vert \varphi _{j}\right\rangle
_{A_{2}B_{2}}\right) ,  \label{pfm1}
\end{gather}%
which is exactly Eq. (\ref{main1}). Substitute $\left\vert \varphi
_{i}\right\rangle _{A_{k}B_{k}}=\left\langle i_{C_{k}}\right\vert \left.
\psi \right\rangle _{A_{k}B_{k}C_{k}}$ into Eq. (\ref{pfm1}) and one will
easily obtain Eq. (\ref{main2}).\hfill $\blacksquare $

 Lemma 2 expresses $M_{ij}$ based on the inner product which is a probability amplitude, so it can not be directly related to the experiment. But a simple change will  immediately arrive at the directly measurable quantities given in our following main theorem.

\textbf{Theorem 1.}-\textit{The absolute value }$\left\vert
M_{ij}\right\vert $\textit{\ can be directly measured by local projective
measurements, i.e.,}%
\begin{equation}
\left\vert M_{ij}\right\vert =2\sqrt{\left\langle \psi \right\vert
_{\varepsilon _{1}}\left\langle \psi \right\vert _{\varepsilon _{2}}\mathcal{%
A}_{ij}\left\vert \psi \right\rangle _{\varepsilon _{1}}\left\vert \psi
\right\rangle _{\varepsilon _{2}}}  \label{fi}
\end{equation}%
\textit{with }$\mathcal{A}$ \textit{a factorizable observable given by} 
\begin{equation}
\mathcal{A}_{ij}\mathcal{=}P_{-}^{(A_{1}A_{2})}\otimes
P_{-}^{(B_{1}B_{2})}\otimes P_{i}^{\left( C_{1}\right) }\otimes
P_{j}^{\left( C_{2}\right) },  \label{ob}
\end{equation}%
\textit{where the subscripts }$\varepsilon _{k}$\textit{\ denote the index }$%
A_{k}B_{k}C_{k}$\textit{\ and }$P_{m}^{\left( C_{k}\right) }=\left\vert
m\right\rangle _{C_{k}}\left\langle m\right\vert $\textit{\ are the
projectors subject to the basis }$\left\vert m\right\rangle $\textit{\ in }$%
H_{C_{k}}$\textit{. Let }$M=U\Lambda U^{T}$, \textit{\ with U unitary and }$%
\Lambda $\textit{\ diagonal and positive. Then }$\tau \left( \left\vert \psi
\right\rangle _{ABC}\right) $\textit{\ can be directly measured via }$%
\left\vert M_{ii}\right\vert $ \textit{with the optimal choice} $P_{i}^{\left( C_{k}\right)
}=U^{*}\left\vert i\right\rangle _{C_{k}}$.

\textbf{Proof. }Eq. (\ref{fi}) is a direct and obvious result of Lemma 2.
This proof is omitted here. Next we will show that the genuine
tripartite entanglement $\tau \left( \left\vert \psi \right\rangle
_{ABC}\right) $ can be obtained by the measurable $\left\vert
M_{ij}\right\vert $. From Ref. \cite{yum}, one can find that any operation $%
Q $ operated on the qudit C can be equivalently described as $\tilde{M}%
=Q^{T}MQ$. Based on Takagi decomposition \cite{horn} of a complex symmetric matrix, one
can always write $M=U\Lambda U^{T}$ where $U$ is a unitary matrix and $%
\Lambda $ is a diagonal matrix with the diagonal entries corresponding to
the singular values of $M$. In this sense, one can always select a proper
local unitary operation $Q$ on qudit C such that $Q^{T}U=I$ which
corresponds to $\tilde{P}_{i}^{\left( C_{1}\right) }=U^{*}\left\vert
i\right\rangle _{C_{1}}$ and $\tilde{P}_{j}^{\left( C_{2}\right)
}=U^{*}\left\vert j\right\rangle _{C_{2}}$. With such a choice, one will
obtain that $\tilde{M}=\left\vert \tilde{M}\right\vert =\Lambda $. In other
words, so long as we choose the optimal projective measurements on C$_{1}$
and C$_{2}$, $\tilde{M}_{ii}$ just corresponds to $\lambda _{i}$, i.e., the
singular values of $M$. This means that $\tau \left( \left\vert \psi
\right\rangle _{ABC}\right) $ can be measured directly and locally based on
Eq. (\ref{3e}).  \hfill $%
\blacksquare $

The above proof implies very important contents. One can find that $\tilde{M}_{ij},$ $i\neq j$, vanish once the
optimal projective measurements on C$_{k}$ are achieved. It means that, if the optimal projector on C$_{1}$ and C$_{2}$ are different,  there will not be any output corresponding to the projective measurements 
$P_{-}^{(A_{1}A_{2})}$ and $P_{-}^{(B_{1}B_{2})}$. So this becomes  an
important index by which one can signal when the optimal measurement basis has been achieved in the practical adjusting procedure.

An intuitive illustration of this scheme is sketched in Fig. 1. Suppose we
have a pair of entangled tripartite pure states $\left\vert \psi
\right\rangle _{A_{1}B_{1}C_{1}}$ and $\left\vert \psi \right\rangle
_{A_{2}B_{2}C_{2}}$. The projective measurements $P_{i}^{\left( C_{1}\right)
}$ and $P_{j}^{\left( C_{2}\right) }$ are performed on qubits C$_{1}$ and C$%
_{2}$, respectively. At the same time, joint projective measurement $%
P_{-}^{(A_{1}A_{2})}$ is performed on A$_{1}$, A$_{2}$ and $%
P_{-}^{(B_{1}B_{2})}$ is performed on B$_{1}$, B$_{2}$. Adjust the
measurement basis of $P_{i}^{\left( C_{1}\right) }$ and $P_{j}^{\left(
C_{2}\right) }$ such that no signal is output from the measurement terminals 
$P_{-}^{(A_{1}A_{2})}$ and $P_{-}^{(B_{1}B_{2})}$ when $i\neq j$. At this
moment, $\tilde{M}_{ii}$ can be expressed by%
\begin{equation}
\tilde{M}%
_{ii}=p_{i}^{(C_{1})}p_{i}^{(C_{2})}p_{-}^{(A_{1}A_{2})}p_{-}^{(B_{1}B_{2})}
\label{ep}
\end{equation}%
with $p_{k}^{(\cdot )}$ denoting the probability corresponding to the
projective measurements on $(\cdot )$. So $\tau \left( \left\vert \psi
\right\rangle _{ABC}\right) $ can be easily obtained.

\section{Measuring 3-tangle of qubits in linear optical experiment}
We take the linear optical experiment of our
scheme as an example. The experimental setup is briefly sketched in Fig. 2.
Two entanglement resources are used to generate three entangled polarized photons with the state $\left\vert \psi \right\rangle
_{A_{1}B_{1}C_{1}} $ and $\left\vert \psi \right\rangle _{A_{2}B_{2}C_{2}}$,
respectively. As a demonstration, one can use the same setups as Ref. \cite{Pan} to
produce the polarized GHZ state of three photons. Each group of entangled
photons are distributed into three paths represented by the labels of the corresponding
qubits, respectively. Let photons $A_{1}$ and $A_{2}$ go through a
beam splitter (BS), then undergo a polarized beam splitter (PBS) and
finally be detected by single photon detectors. At the same time, let photons $%
B_{1}$ and $B_{2}$ go through another set of  similar setups as $A_{1}$ and $%
A_{2}$. But we let photons $C_{1}$ and $C_{2}$ first go through a quarter wave
plate (QWP) and a half wave plate (HWP) and then arrive at a PBS. One can practically
adjust the QWP and HWP to look for the optimal basis for projective
measurements $P_{i}^{\left( C_{1}\right) }$ and $P_{j}^{\left( C_{2}\right) }
$ which can be achieved until no effective click \cite{note} corresponding to $%
P_{-}^{(A_{1}A_{2})}$ and $P_{-}^{(B_{1}B_{2})}$ is recorded. The final
measurement statistics can be achieved by recording the clicks of each
detector. It is worthwhile  to note that such an adjustment of basis has
been employed in a previous experiment \cite{xu}, and hence it is not at all necessary to worry
about the feasibility of the choice of the optimal basis. 
\begin{figure}[tbp]
\includegraphics[width=8.5cm]{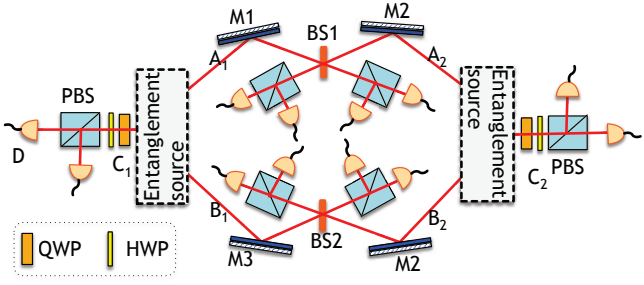}
\caption{(Color online) Brief diagram for the linear optical realization. Three photons are distributed to three paths. The BS+PBS is used to realize the projective measurement $P_-^{(\cdot)}$ and the HWP+QWP is 
used to  implement the projective measurement $P_{C_k}$ on any possible basis. The entanglement sources can be replaced by the experiment in Ref. \cite{Pan} for a simple demonstration of the GHZ state.}
\label{2}
\end{figure}

\section{The measurable lower bound}
Even though the adjustment of projective
measurements on qubit $C_{k}$ is practically feasible, one could not be
satisfied with this. Next, we present a weak scheme that does not need the
optimal $P_{-}^{(A_{1}A_{2})}$ and $P_{-}^{(B_{1}B_{2})}$. So this will
greatly simplify the practical operations, but the cost is that one needs more measurement outcomes and  only a lower bound of $\tau \left( \left\vert \psi \right\rangle
_{ABC}\right) $ is obtained, even though the bounds are very good.

From Eq. (\ref{3e}), one can easily find that 
\begin{eqnarray}
\tau \left( \left\vert \psi \right\rangle _{ABC}\right)  &=&\left\{ 
\begin{array}{cc}
\sum\limits_{i=1}^{4}\lambda _{i}, & \lambda _{1}\leq
\sum\limits_{i=2}^{4}\lambda _{i} \\ 
2\sqrt{\lambda _{1}\sum\limits_{i=2}^{4}\lambda _{i}} & \lambda
_{1}>\sum\limits_{i=2}^{4}\lambda _{i}%
\end{array}%
\right.   \notag \\
&\geqslant &2\sqrt{\sum\limits_{i=1}^{4}\lambda _{i}^{2}-\lambda _{1}^{2}}.
\label{lb1}
\end{eqnarray}%
This is a good lower bound in that it is a sufficient and necessary
condition for GHZ type inseparability of $\left\vert \psi \right\rangle _{ABC}
$.  It can be seen from that the lower bound vanishes if and only if the matrix $M$ is rank-one which
is equivalent to $\tau$. Based on the upper bound of the singular value of a matrix \cite{horn}, one can find that $\lambda _{1}$ can be well bounded
by%
\begin{gather}
\lambda _{1}\leq \sigma _{U}\left( q\right)   \notag \\
=\left[ \max_{k}\sum\limits_{j=1}^{n}\left\vert M_{jk}\right\vert ^{2q}%
\right] ^{1/2}\left[ \max_{j}\sum\limits_{k=1}^{n}\left\vert
M_{jk}\right\vert ^{2(1-q)}\right] ^{1/2}  \label{gs}
\end{gather}%
for $q\in \lbrack 0,1]$. Thus we have%
\begin{equation}
\tau\left( \left\vert \psi \right\rangle _{ABC}\right) \geqslant 2\sqrt{%
TrMM^{\dag }-\min_{q\in \lbrack 0,1]}\sigma _{U}^{2}\left( q\right) }.
\label{lba}
\end{equation}%
Some simple bounds can be found when $q=0,\frac{1}{2},1$.

It is obvious that $TrMM^{\dag }=Tr\rho _{AB}\left( \sigma _{y}\otimes
\sigma _{y}\right) \rho _{AB}^{\ast }\left( \sigma _{y}\otimes \sigma
_{y}\right) =Tr\left[ \left( \rho _{A_{1}B_{1}}\otimes \rho
_{A_{2}B_{2}}\right) P_{-}^{(A_{1}A_{2})}\otimes P_{-}^{(B_{1}B_{2})}\right] 
$ which shows that $TrMM^{\dag }$ can be directly measured by local
projective measurements with a two-fold copy of the state. In addition, $%
\sigma _{U}\left( q\right) $ given in Eq. (\ref{gs}) is described by $\left\vert M_{jk}\right\vert $, which can be
obtained by the measurement statistics produced by $n(n+1)/2$ measurements
including $P_{i}^{\left( C_{1}\right) }$ and $P_{j}^{\left( C_{2}\right) }$.
From the lower bound point of view, it is not necessary to adjust the basis
for $P_{i}^{\left( C_{1}\right) }$ and $P_{j}^{\left( C_{2}\right) }$.  But
the optimal choice of these two projectors can greatly improve the lower
bound. So  the lower bound is locally measurable provided that two
copies of the states are available.

In fact, for three qubits, an alternative lower bound that could  be relatively tight can be given by
\begin{eqnarray}
\tau_{qubit} \left( \left\vert \psi \right\rangle _{ABC}\right) &=& 2\sqrt{\left\vert\det{M}\right\vert} \notag\\
&\geq& 2\sqrt{\left\vert\left\vert M_{00}\right\vert\left\vert M_{11}\right\vert-\left\vert M_{01}\right\vert\left\vert M_{10}\right\vert\right\vert}.
\label{lba}
\end{eqnarray}%
This bound can be easily proved by Eq. (\ref{mm}) for $ i,j\leq 1$. So Eq. (\ref{3e}) can be directly related to the determinant of matrix $M$. It is obvious that all of the elements in the lower bound can be directly measured based on the above scheme, so the lower bound can be experimentally determined.

\section{Discussions and conclusion}
No experiment is perfect, so we have to
know to what degree the measurement results are acceptable. In this scheme, in
order to reduce the copies of the measured state, a key point is to adjust
the projective measurements on C$_{k}$ such that no effective outputs
corresponding to $P_{-}^{(A_{1}A_{2})}\otimes P_{-}^{(B_{1}B_{2})}$ are
generated. However, there could be a small probability $\epsilon $ to detect
photons in a practical scenario. Thus, the practical $\left\vert
M_{ii}\right\vert $ can always be formally given by $\left\vert
M_{ii}\right\vert =\lambda _{i}+\epsilon \Delta $ in first order. This will
lead to a small ($\backsim\epsilon $) deviation for the exact tripartite
entanglement, but a little smaller lower bound for Eq. (\ref{lba}). In addition,
the previous similar jobs tried to compensate for the experimental
imperfection, which could imply that more prior information should be known.
Here instead of doing this, we will mainly find out the potential errors. Without loss of generality, we only suppose that the prepared state is
a quasi-pure state, that is, 
\begin{equation}
\rho _{A_{k}B_{k}C_{k}}=(1-\epsilon _{k})\left\vert \psi \right\rangle
\left\langle \psi \right\vert _{A_{k}B_{k}C_{k}}+\epsilon _{k}\varrho _{k},\label{im1}
\end{equation}%
where $\epsilon _{k}\ll 1$ and $\varrho _{k}$ is a general tripartite
density matrix with the subscript $k=1,2$ distinguishing different copies.
Since the copy of the state is generated by another setup, it is reasonable
to assume that 
\begin{equation}
\left\vert \psi \right\rangle _{A_{2}B_{2}C_{2}}=\sqrt{1-\epsilon_0 ^{2}}%
\left\vert \psi \right\rangle _{A_{1}B_{1}C_{1}}+\epsilon _{0}\left\vert
\phi \right\rangle \label{im2}
\end{equation}%
with $\epsilon _{0}\ll 1$ and $\left\langle \psi \right\vert \left. \phi
\right\rangle =0$. Substitute Eqs. (\ref{im1}) and (\ref{im2}) into Eq. (\ref{ob}), and one will find that $%
\left\vert M_{ij}^{\prime }\right\vert $, corresponding to the imperfect
preparation and copy, can be written in first order of $\epsilon _{i}$ as%
\begin{equation*}
\left\vert M_{ij}^{\prime }\right\vert \backsim \sqrt{(1-\epsilon
_{1}-\epsilon _{2})\left\vert M_{ij}\right\vert ^{2}+\tilde{\epsilon}N},\label{imr}
\end{equation*}%
where $\tilde{\epsilon}=\max \{\epsilon _{0},\epsilon _{1},\epsilon _{2}\}$
and $N$ is not explicitly given here. All of the above analysis shows that the
imperfect experiment will lead to a small deviation (about $\max\{\epsilon,\tilde{\epsilon}\}$) of the real value. However,
if the entanglement of $\left\vert\psi\right\rangle$ is so small that $\tau\backsim\epsilon$, that is, noise drowns out the signal, then this scheme cannot detect any entanglement which is similar to all the relevant jobs.

In summary, we have found that the high-dimensional tripartite entanglement
can be locally measured with only a two-fold copy of the state. For simplicity,
we also provide a good lower bound for entanglement which will simplify the
practical operations but require more measurement outcomes. As a demonstration, we
consider how to measure the three-tangle for three entangled qubits based on
linear optical setups.  The current scheme is only fit for a pure state which
may not be so practical. However, needless to say, for the measurable
entanglement with less copies of the state, even a simple lower bound for
tripartite entanglement is not available in entanglement theory. We think this
scheme could be an important step towards the more general cases. 

\section{Acknowledgements}
This work was supported by the National Natural
Science Foundation of China, under Grants No.11375036 and No. 11175033, the
Xinghai Scholar Cultivation Plan, and the Fundamental Research Funds for the
Central Universities under Grants No. DUT15LK35 and No. DUT15TD47.


\begin{thebibliography}{99}
\bibitem{note} Ideally, no click should be recorded. However, every experiment has its tolerable error $\varepsilon$. 
Within such an error, a few clicks (ineffective clicks) should be allowed for simplifying the practical operations.
\end{thebibliography}
\end{document}